\documentclass{article}
\usepackage{spconf, amsmath, graphicx}
\usepackage{xcolor, aligned-overset}
\usepackage{booktabs, adjustbox, multirow, multicol, amsfonts, url, enumitem}

\usepackage[hang]{footmisc}

\def\snake{{\includegraphics[width=0.02\textwidth]{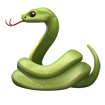}}{}}

\title{Speech Slytherin: Examining the Performance and Efficiency of Mamba for Speech Separation, Recognition, and Synthesis}
\name{Xilin Jiang\textsuperscript{\snake}, Yinghao Aaron Li\textsuperscript{\snake\thanks{\snake Equal contribution.}}, Adrian Nicolas Florea, Cong Han, Nima Mesgarani}
\address{Department of Electrical Engineering, Columbia University, USA}

\usepackage{hyperref}

\begin{document}
%
\maketitle
\begin{abstract}
It is too early to conclude that Mamba is a better alternative to transformers for speech before comparing Mamba with transformers in terms of both performance and efficiency in multiple speech-related tasks. To reach this conclusion, we propose and evaluate three models for three tasks: Mamba-TasNet for speech separation, ConMamba for speech recognition, and VALL-M for speech synthesis. We compare them with transformers of similar sizes in performance, memory, and speed. Our Mamba or Mamba-transformer hybrid models show comparable or higher performance than their transformer counterparts: Sepformer, Conformer, and VALL-E. They are more efficient than transformers in memory and speed for speech longer than a threshold duration, inversely related to the resolution of a speech token. Mamba for separation is the most efficient, and Mamba for recognition is the least. Further, we show that Mamba is not more efficient than transformer for speech shorter than the threshold duration and performs worse in models that require joint modeling of text and speech, such as cross or masked attention of two inputs. Therefore, we argue that the superiority of Mamba or transformer depends on particular problems and models. Code available\footnote{Mamba-TasNet: https://github.com/xi-j/Mamba-TasNet \\
ConMamba: https://github.com/xi-j/Mamba-ASR
}.
\end{abstract}
\section{Introduction}
\label{sec:intro}
Speech and text are two closely related modalities, and both can be a long sequence of hundreds or thousands of tokens. Both the local and context information are necessary to understand either one modality alone or to translate between them. Therefore, a powerful and efficient sequence modeling mechanism, usually a recurrent neural network (RNN) \cite{lstm} or a transformer \cite{transformer}, is commonly used in the literature to comprehend both local nuances and global context. In particular, transformer models have become the de facto choice for many popular speech tasks, including separation, recognition, and synthesis, due to the reliability and scalability of performance. However, efficiency-wise, transformer models suffer from quadratic complexity in token length, which is unfriendly for long speech and text sequences. A good tradeoff between performance and efficiency needs to be found for model deployment with memory and time constraints. 

\begin{figure}[!t]
  \centering  \includegraphics[width=0.95\columnwidth]{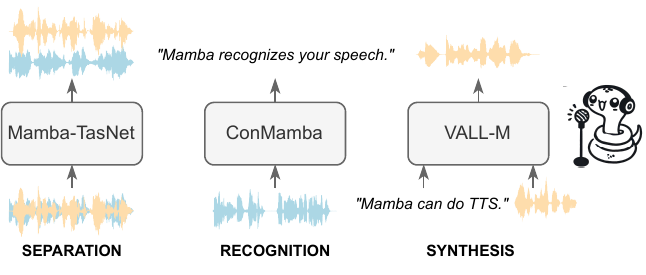}
  \caption{We propose three models for speech separation, recognition, and synthesis.}
  \label{fig:3mamba}
\end{figure}

A recently proposed state space model, Mamba, appears to be a strong candidate for sequence modeling tasks \cite{mamba}. Mamba enjoys linear complexity in token length and performs on par with transformers. This performance is first demonstrated in text, image, and biomedical data \cite{mamba, bimamba}. Several recent works have also applied Mamba to audio and speech \cite{mamba_se1, dpmamba, audio_mamba1}. However, despite the popularity of Mamba, most research focuses on a single task and emphasizes performance. To fully investigate if Mamba is a better alternative to transformers, it is essential to compare it across multiple tasks and evaluate its memory and speed efficiency as well.

To complete this puzzle, we make a thorough performance and efficiency comparison between Mamba and transformers for three speech and text modeling scenarios in this paper: speech-to-speech, speech-to-text, and text-to-speech. We study a representative task for each scenario: speech separation, automatic speech recognition (ASR), and text-to-speech synthesis (TTS). Respectively, we propose and evaluate Mamba-TasNet for separation, ConMamba for recognition, and VALL-M and VALL-ME for synthesis in rivals of state-of-the-art transformer models Sepformer, Conformer, and VALL-E. These models differ in both functionality and architecture, particularly in terms of the resolution of each speech token and, therefore, the number of tokens for a continuous speech. Based on our performance analysis and memory and speed benchmark for speech inputs or outputs of different durations, we make observations below:
\begin{itemize}[noitemsep]
 \item Bidirectional Mamba performs similarly or better than self-attention in Mamba-TasNet and the ConMamba encoder than Sepformer and the Conformer encoder for separation and ASR.
 \item Unidirectional Mamba performs slightly worse than masked or cross-attention to joint text and speech inputs in the ASR decoder and VALL-E's autoregressive language model. A hybrid Mamba encoder and transformer decoder model performs similarly or better than transformer encoder-decoder models for ASR and TTS.
 \item Mamba becomes significantly more efficient in memory and speed for speech longer than a threshold duration. The threshold depends on the resolution of a speech token. Mamba has a greater advantage in high-resolution tasks such as separation but a minor or no advantage in low-resolution tasks such as ASR.
\end{itemize}
We hope our observations can encourage consideration of the more suitable use cases of Mamba in speech and promote better and more efficient model designs.

\section{Related Works}
\label{sec:literature}

Mamba has shown a transformer-level performance in numerous modalities that can represented as a sequence, including text \cite{mamba, mamba_llm}, images \cite{bimamba,  xu2024survey}, videos \cite{yang2024vivim, chen2024video}, and biomedical data \cite{mamba, ma2024u}. Audio or speech, either in waveform or spectrogram, is naturally a sequence. Most early works apply Mamba to a single audio or speech task:
\cite{mamba_se1, mamba_se2, mamba_se3} to speech enhancement,  \cite{dpmamba, spmamba} to speech separation, and \cite{chen2024rawbmamba, lin2024audio} to audio detection and classification. Other works \cite{audio_mamba1, audio_mamba2, audio_mamba3} propose a self-supervised audio transformer trained with masked spectrogram modeling. \cite{mamba_in_speech, Miyazaki2024mamba} are the two most comprehensive studies of applications of Mamba in speech at the time this paper is written. \cite{mamba_in_speech} studies Mamba for speech enhancement and recognition. \cite{Miyazaki2024mamba} studies Mamba for speech recognition, synthesis, understanding, and summarization. Despite the wide range of tasks, they focus on performance and lack speed and memory comparison with respect to speech duration between Mamba and transformers. Therefore, a solid conclusion that Mamba is better than transformers for speech in all scenarios is yet to be made. This work does not search for a particular scenario where Mamba is better than transformers. Instead, we make a fair comparison between Mamba and transformer in multiple identical settings.

\begin{figure}[!t]
  \centering  \includegraphics[width=0.9\columnwidth]{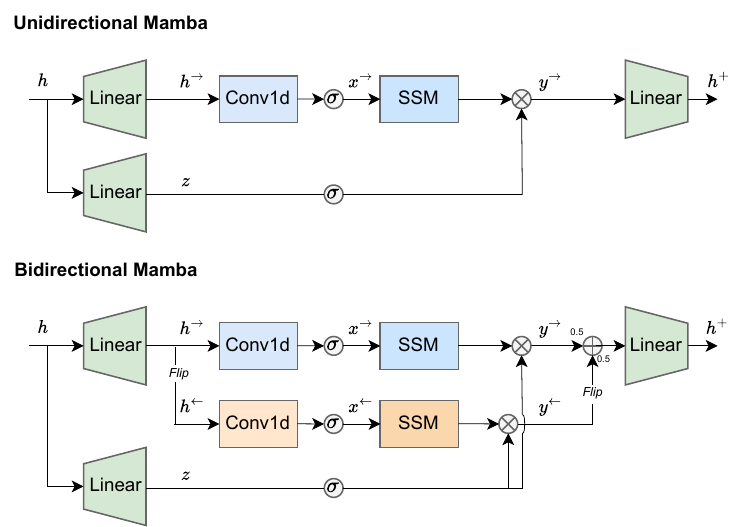}
  \caption{The architecture of unidirectional and bidirectional Mamba.}
  \label{fig:mamba}
\end{figure}

\begin{figure*}[!ht]
  \centering  
  \includegraphics[width=0.88\textwidth]{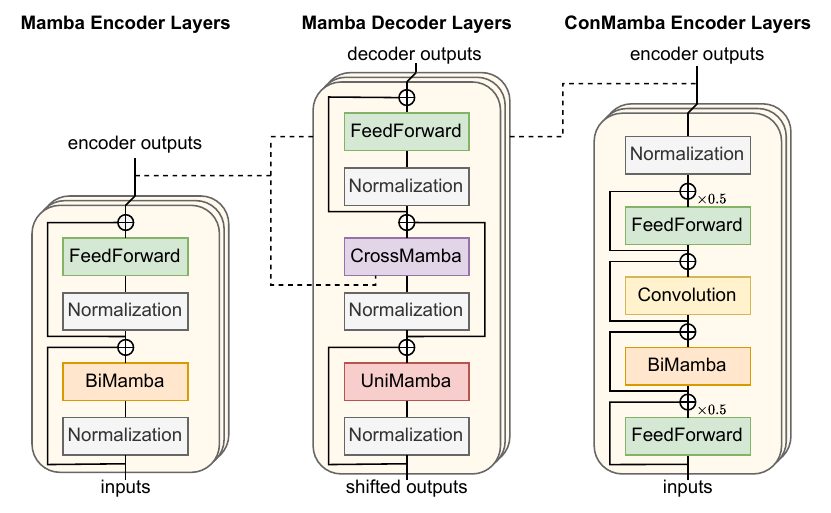}
  \caption{Mamba encoder, decoder, and ConMamba encoder layers. When used alone, the FeedForward module in the encoder and decoder is optional (none in Mamba-TasNet), and the CrossMamba module in the decoder is not needed.}
  \label{fig:block}
\end{figure*}

\section{Methods}
\label{sec:methods}
We present our Mamba models in a bottom-up manner. We first introduce the shared building blocks of all models in Sec.\ref{prelim} and then move to particular model designs in Sec.\ref{models}.

\subsection{Preliminary} \label{prelim}

\subsubsection{Unidirectional and Bidirectional Mamba}

The core of Mamba is a linear selective state space model (SSM). In the equation below, $\textbf{h}_t$, $\textbf{x}_t$ and $\textbf{y}_t$ are the state, input, and output at time $t$. \textbf{A}, \textbf{B}, and \textbf{C} are learnable parameters corresponding to the state transition matrix, input matrix, and output matrix. 
\begin{align}
    \textbf{h}_t = \textbf{A}\textbf{h}_{t-1} + \textbf{B} \textbf{x}_t, \hspace{4pt} \textbf{y}_t = \textbf{C}\textbf{h}_t \label{eq:discrete}
\end{align}
Thanks to its linearity, we can express the entire sequence \textbf{y} of length $L$ as a convolution between \textbf{x} of the same length and a kernel \textbf{K}. Since $\textbf{A}_t$, $\textbf{B}_t$, and $\textbf{C}_t$ are all dependent on input $\textbf{x}_t$ (i.e. selective), Equation \ref{eq:kernel} cannot be computed directly but is computed by a parallel scan algorithm instead.
\begin{align}
    \textbf{K} = (C\textbf{B}, \textbf{A}\textbf{B}, ..., C\textbf{A}^{L-1}\textbf{B}), \hspace{4pt} \textbf{y} = \textbf{x} \ast \textbf{K} \label{eq:kernel}
\end{align}
A (unidirectional) Mamba is an SSM sandwiched by gated linear layers \cite{mamba}. Mamba is non-linear but causal. For most speech-related tasks, such as speech separation and recognition, bidirectional modeling is prefered since it allows for the incorporation of both past and future information. Therefore, we borrow the bidirectional Mamba proposed in \cite{bimamba} for these non-causal tasks. The architectures of both unidirectional and bidirectional Mamba are illustrated in Fig.\ref{fig:mamba}. For bidirectional Mamba, two SSMs and causal convolutions run in parallel, one for the original sequence and the other one for the reversed sequence. The outputs of the SSMs are averaged to incorporate information in both directions. 

\subsubsection{Mamba Encoder and Decoder}
Common add-ons to stabilize training and improve performance for attention, including feature normalization and residual connection, can also be applied to Mamba. Following the naming convention of transformer \cite{transformer}, we call a block with bidirectional Mamba, normalization, and residual connection a \textit{Mamba encoder layer} and a block with unidirectional Mamba a \textit{Mamba decoder layer}. We present these layers in Fig.\ref{fig:block}, along with a \textit{ConMamba encoder layer}, which we will elaborate on in the next section. Notice that we can optionally add a feedforward module after the Mamba module. This design has been justified in Mamba-based large language models \cite{mamba_llm, jamba}. We follow this design to build codec language models for VALL-M and VALL-ME.

Encoder-decoder models, a large school of transformers, have both encoder and decoder layers. Their features are merged together by cross-attention, with the encoder outputs as the keys ($k$) and values ($v=k$) and the decoder features as the queries ($q$). $k$ and $q$ are often of different lengths or even modalities (e.g. speech and text). Unfortunately, there is no native analog of cross-attention in Mamba that can handle multiple inputs of variable lengths. To address this, we propose \textit{CrossMamba}, a unidirectional Mamba on two or more inputs, as a plug-in replacement for cross-attention. We simply concatenate $k$ and $q$ together as the inputs and only keep the latter half of the outputs with the same length as the query:
\begin{align}
    \textit{CrossMamba}(k, q) = \textit{UniMamba}(\textit{cat}(k, q))[-len(q):]
\end{align}
More general, more than two inputs $\{x\}_n = x_1, x_2, ..., x_n$ can be processed in the same way:
\begin{align}
    \textit{CrossMamba}(\{x\}_n ) = \textit{UniMamba}(\textit{cat}(\{x\}_n ))[-len(x_n):]
\end{align}
\subsection{Models} \label{models}

\subsubsection{Mamba-TasNet}

Mamba-TasNet follows the design of Conv-TasNet \cite{convtasnet} with a linear waveform encoder \& decoder pair and a mask estimation network (MaskNet) in between. The full architecture is drawn in Fig.\ref{fig:3mamba}. The MaskNet is a Mamba encoder model. A stack of Mamba encoder blocks processes the encoded features of the mixture. $S$ masks corresponding to $S$ sources are estimated from the features after the last block. The separated sources are reconstructed to waveforms by the decoder.

Notice that Mamba-TasNet is a \textit{single-path} model. Time-domain RNN or transformer-based separation models after DPRNN \cite{dprnn} often follow a dual-path (or multi-path) architecture for both performance and efficiency considerations \cite{dptransformer, sepformer, mossformer}. An earlier Mamba separation model, DPMamba \cite{dpmamba}, also adopts a dual-path architecture. In these models, a waveform (downsampled by the encoder but still thousands of tokens) is split into chunks of the same sizes. Intra-chunk and inter-chunk RNNs or transformers each process a sequence much shorter in length than the waveform to alleviate the high complexity of RNN and transformer in sequence length. On the other hand, since Mamba enjoys linear complexity in sequence length, we hypothesize that dual-path architecture is not necessary for Mamba. Our results prove this hypothesis.

\begin{figure*}[!ht]
  \centering  
  \includegraphics[width=0.9\textwidth]{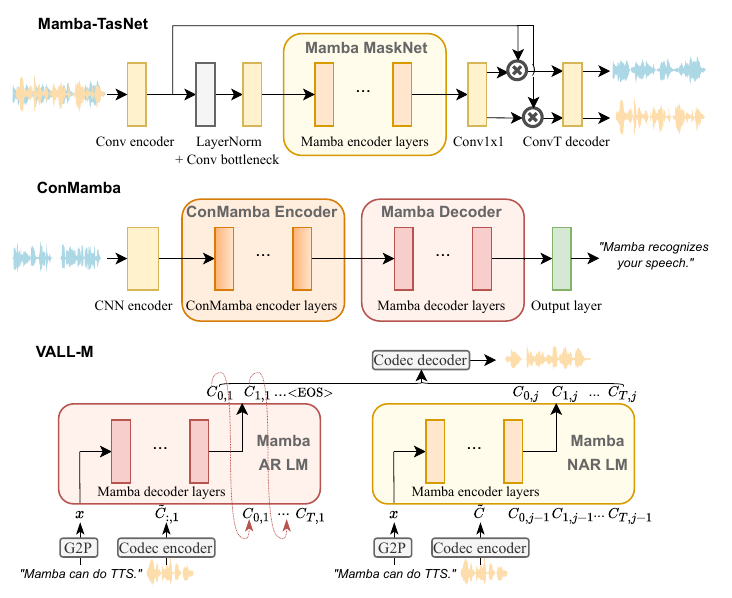}
  \caption{The architecture of Mamba-TasNet, ConMamba, and VALL-M.}
  \label{fig:models}
\end{figure*}

\subsubsection{ConMamba}
Conformer \cite{conformer} proposes to add a convolution module in the transformer encoder block to effectively exploit local features. This inspires us to add a convolution module to the Mamba encoder block. The new ConMamba block is depicted at the bottom of Fig.\ref{fig:block}. A ConMamba block contains Mamba, feedforward, and convolution modules. The computation workflow from input $x$ to output $y$ is followed:
\setlength{\abovedisplayskip}{3pt}
\setlength{\belowdisplayskip}{3pt}
\begin{align}
    & x^{\prime}= x + \frac{1}{2} \textit{FeedForward}(x) \\
    & x^{\prime\prime} = x^{\prime} + \textit{BiMamba}(x^{\prime}) \\
    & x^{\prime\prime\prime} = x^{\prime\prime} + \textit{Convolution}(x^{\prime\prime}) \\
    & y = \textit{LayerNorm}(x^{\prime\prime\prime} + \frac{1}{2} \textit{FeedForward}(x^{\prime\prime\prime}))
\end{align}
\setlength{\abovedisplayskip}{\baselineskip}
\setlength{\belowdisplayskip}{\baselineskip}
ConMamba's encoder consists of a stack of ConMamba encoder blocks following a CNN frontend to compress the spectrogram into tokens, as illustrated in the middle of Fig.\ref{fig:3mamba}. Then, we can estimate text token probability from the last encoder representation or add a transformer or Mamba decoder to output tokens autoregressively. \cite{mamba_in_speech} has implemented the latter approach with a Mamba encoder and a transformer decoder and reported a higher performance than the Conformer. However, since half of their model is still transformer, whether Mamba alone contributes to the high performance is still to be determined. To make our results more convincing, we implement and compare a ConMamba encoder-only, ConMamba encoder-Mamba decoder, and ConMamba encoder-transformer decoder model against a Conformer encoder-only and Conformer encoder-transformer decoder model.

\subsubsection{VALL-M}
VALL-E \cite{valle} formulates speech synthesis as a language modeling task in a speech codec. The tokens in this language are the codes from 8 hierarchical codebooks in EnCodec \cite{encodec}. The language modeling task aims to maximize the conditional probability $p(\mathbf{C}|\mathbf{x}, \tilde{\mathbf{C}}; \theta)$, where $\mathbf{x}$ is the phoneme sequence of the speech to synthesize, $\tilde{\mathbf{C}} \in \mathbb{R}^{\tilde{T} \times 8}$ is 8 sequences of codes tokenized from a short enrollment of the speaker to clone, $\mathbf{C} \in \mathbb{R}^{T \times 8}$ is the code sequences of the synthesized speech, and $\theta = \{\theta_{\text{AR}}, \theta_{\text{NAR}}\}$ denotes the model parameters. $p(\mathbf{C}|\mathbf{x}, \tilde{\mathbf{C}}; \theta)$ can be optimized by solving two separate language modeling tasks due to the hierarchy of codes. An autoregressive (AR) task maximizes the next code probability of the code sequence obtained from the first quantizer:
\setlength{\abovedisplayskip}{0pt}
\setlength{\belowdisplayskip}{0pt}
\begin{align}
    p(C_{:,1}|\mathbf{x}, \tilde{\mathbf{C}}_{:,1}; \theta_{AR}) = \prod_{t=0}^{T} p(C_{t,1}|C_{<t,1}, \tilde{\mathbf{C}}_{:,1}, \mathbf{x}; \theta_{AR}) \label{eq:ar}
\end{align}
\setlength{\abovedisplayskip}{\baselineskip}
\setlength{\belowdisplayskip}{\baselineskip}
The other non-autoregressive (NAR) task works on details from the second to the last residual codebooks. It is allowed to look at the entire code sequence from the first quantizer:
\setlength{\abovedisplayskip}{\baselineskip}
\setlength{\belowdisplayskip}{\baselineskip}
\begin{align}
    p(\mathbf{C}_{:,2:8}|\mathbf{x}, \tilde{\mathbf{C}}; \theta_{NAR}) = \prod_{j=2}^{8} p(C_{:,j}|\mathbf{C}_{:,<j}, \mathbf{x}, \tilde{\mathbf{C}}; \theta_{NAR})
    \label{eq:nar}
\end{align}
\setlength{\abovedisplayskip}{\baselineskip}
\setlength{\belowdisplayskip}{\baselineskip}
Due to the causal (for speech tokens) or non-causal nature of the tasks, we solve the AR task with a Mamba decoder and the NAR task with a Mamba encoder model. We call this model VALL-M. We provide another model with a transformer decoder model for the AR task and a Mamba encoder model for the NAR task. We call this hybrid model VALL-ME.

\begin{table}[!t]
    \caption{Model configuration of all Mamba and transformer models. + 4 and + 6 means 4 or 6 additional Mamba or transformer decoder layers for encoder-decoder ASR. 12 + 12 means 12 layers for both AR and NAR language models for VALL-E, VALL-M, and VALL-ME. \\}
    \begin{adjustbox}{width=\columnwidth,center}
    \begin{tabular}{c|cccc}
    \toprule
    Model  & Dimension $D$  & \# Layers & Token Res (ms) & \#Tokens for 10s \\ \hline \hline
    Sepformer & 256 & 16 $\times$ 2  & 1 & 10,000 \\
    Mamba-TasNet (M) & 256 & 32   & 1 & 10,000 \\
    Mamba-TasNet (L) & 512 & 32   & 1 & 10,000 \\
    \midrule
    Conformer (S) & \multirow{2}{*}{144} & \multirow{2}{*}{12 + 4} & \multirow{2}{*}{40} & \multirow{2}{*}{250} \\
    ConMamba (S) &  &  &  & \\
    Conformer (L) & \multirow{2}{*}{512} &\multirow{2}{*}{12 + 6} & \multirow{2}{*}{40} & \multirow{2}{*}{250} \\
    ConMamba (L) & & & & \\
    Conformer (CTC) & \multirow{2}{*}{256} & \multirow{2}{*}{18} & \multirow{2}{*}{40} & \multirow{2}{*}{250} \\
    ConMamba (CTC) & &  & & \\
    \midrule
    VALL-E  & \multirow{3}{*}{1024} & \multirow{3}{*}{12 + 12} & \multirow{3}{*}{13 $\frac{1}{3}$} & \multirow{3}{*}{750} \\
    VALL-M  & & & & \\
    VALL-ME  & & & & \\
    \bottomrule
    \end{tabular}
    \end{adjustbox}
    \label{table:config}
\end{table}

\begin{figure*}[!t]
  \centering  
  \includegraphics[width=0.99\textwidth]{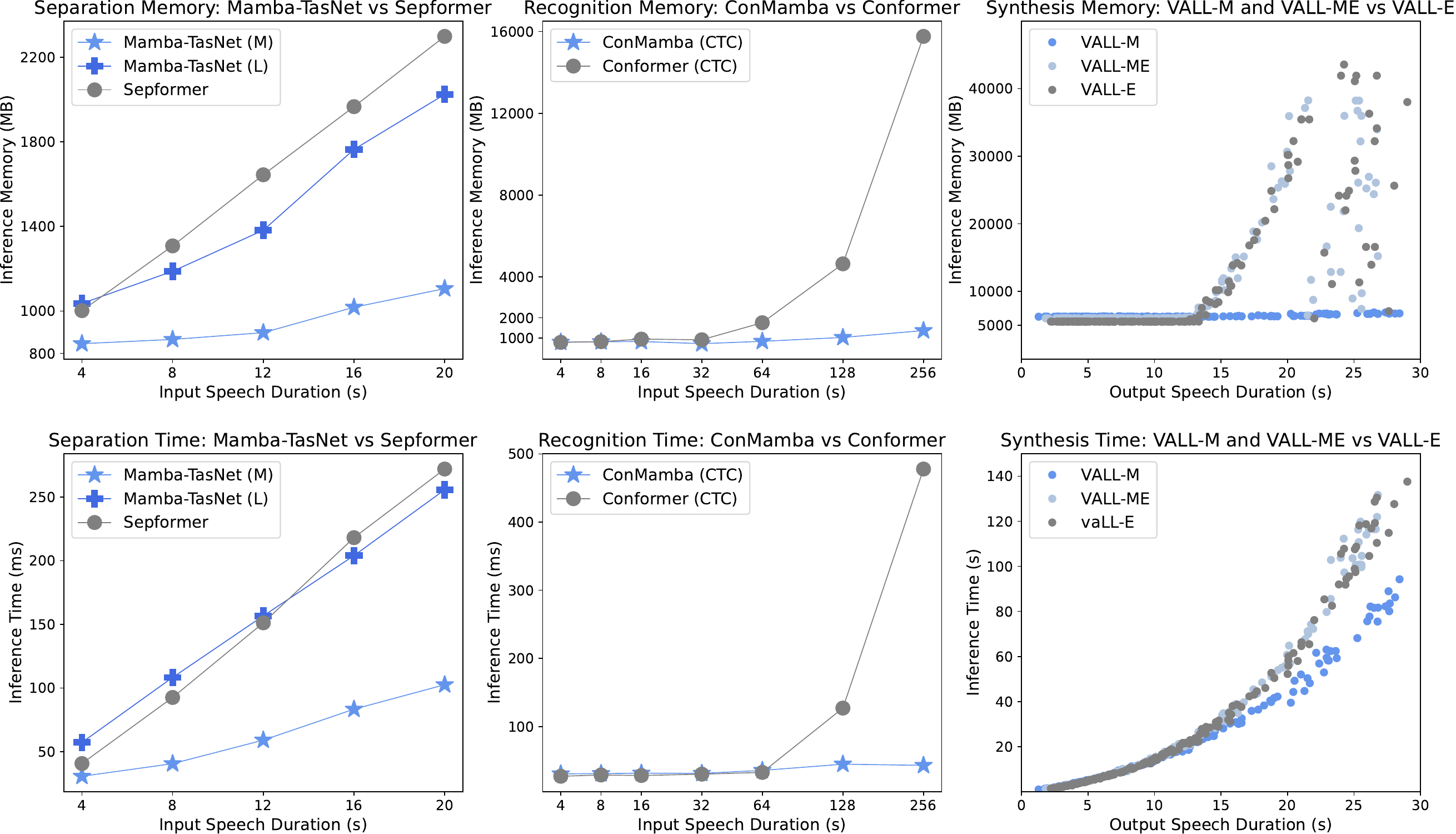}
  \caption{Comparisons of memory and speed for transformer and Mamba models in speech separation, recognition, and synthesis, with respect to speech inputs or outputs of different durations, benchmarked in an NVIDIA L40 GPU with 48GB memory.}
  \label{fig:benchmark}
\end{figure*}

\begin{table}[t]
    \caption{Speech separation signal quality of Mamba-TasNet, DPMamba, and Sepformer on WSJ0-2mix test set.\\}
    \begin{adjustbox}{width=0.9\columnwidth,center}
    \begin{tabular}{c|cc|cc}
    \toprule
    Model  & SI-SNRi (dB) & SDRi (dB) & \#Params (M) \\ \hline \hline
    Sepformer \cite{sepformer} & 22.3 & 22.4  & 25.7   \\
    DPMamba (M) \cite{dpmamba} & \textbf{22.6} & \textbf{22.7} & 15.9  \\
    \textbf{Mamba-TasNet} (M) & 22.4 & 22.6  & 15.6  \\ \hline
    QDPN \cite{QDPN} & 23.6 & n.r.  & 200  \\
    DPMamba (L) & 23.4 & 23.6  & 59.8 \\
    \textbf{Mamba-TasNet} (L) & \textbf{23.7} & \textbf{23.8}  & 59.6 \\ \bottomrule
    \end{tabular}
    \end{adjustbox}
    \label{table:ss}
    \hspace{-5pt}
\end{table}

\begin{table}[t]
    \caption{Speech recognition WER (\%) of ConMamba and Conformer on LibriSpeech test sets.\\}
    \begin{adjustbox}{width=\columnwidth,center}
    \begin{tabular}{c|cc|cc|c}
    \toprule
    Model & Encoder  & Decoder & test-clean  & test-other &\#Params (M)  \\ \hline \hline 
    \multicolumn{6}{c}{\textit{without LM}} \\
    \hline
    Conformer (S) \cite{conformer} & Conformer & Transformer & 4.1 & 10.0 & 13.3    \\
    \multirow{2}{*}{\textbf{ConMamba} (S)} & ConMamba & Transformer & \textbf{4.0} & \textbf{9.5} & 14.1  \\
     & ConMamba & Mamba & \textbf{4.0} & 9.7 & 15.0 \\
    \hline
    Conformer (L) & Conformer & Transformer & \textbf{2.6} & \textbf{6.7} & 109.1  \\
    \multirow{2}{*}{\textbf{ConMamba} (L)} & ConMamba & Transformer & 2.8 & \textbf{6.7} & 115.2 \\ 
    & ConMamba & Mamba & 3.0 & 7.0 & 122.9 \\ 
    \hline
    Conformer (CTC) & \multicolumn{2}{c|}{Conformer Encoder Only} & 4.3 & 11.3 & 28.8  \\
    \textbf{ConMamba} (CTC) &\multicolumn{2}{c|}{ConMamba Encoder Only}  & \textbf{3.9} & \textbf{10.3} & 31.6  \\ \hline 
    \multicolumn{6}{c}{\textit{with LM}} \\
    \hline
    Conformer (S) & Conformer & Transformer & 2.5 & 6.1 & 13.3   \\
    \multirow{2}{*}{\textbf{ConMamba} (S)} & ConMamba & Transformer & \textbf{2.4} & \textbf{5.8} & 14.1  \\
     & ConMamba & Mamba & 2.5 & 6.5 & 15.0  \\
    \hline
    Conformer (L) & Conformer & Transformer & \textbf{2.0} & \textbf{4.5} & 109.1  \\
    \multirow{2}{*}{\textbf{ConMamba} (L)} & ConMamba & Transformer & 2.1 & 4.9 & 115.2 \\ 
    & ConMamba & Mamba & 2.4 & 5.7 & 122.9 \\ 
    \bottomrule
    \end{tabular}
    \end{adjustbox}
    \label{table:asr}
\end{table}

\begin{table}[t]
    \caption{Speech synthesis subjective evaluation of VALL-M and VALL-ME against VALL-E. \\}
    \begin{adjustbox}{width=\columnwidth,center}
    \begin{tabular}{c|cc|cc|c}
    \toprule
    Model  & AR & NAR & CMOS-N & CMOS-S & \#Params (M) \\ \hline \hline
    VALL-E \cite{valle} & Transformer & Transformer  & \textbf{0.00} & 0.00  & 367.5  \\
    \textbf{VALL-M} & Mamba & Mamba & -0.41 & -0.25 & 431.1   \\
    \textbf{VALL-ME} & Transformer & Mamba & -0.01 & \textbf{0.20} & 401.5    \\
    \bottomrule
    \end{tabular}
    \end{adjustbox}
    \label{table:tts}
\end{table}

\section{Results}
\label{sec:results}

\subsection{Models}
\label{subsec:models}

We implemented Mamba-TasNet, ConMamba, and VALL-M of the same dimension and number of layers as their transformer counterparts, as documented in Table.\ref{table:config}. Token resolution refers to the duration of a speech token that makes up the input sequence to the transformer or Mamba layers, equal to the total downsampling stride of the feature extractor and CNN frontend divided by the sampling rate. Separation models have the smallest token resolution of 1 ms to reconstruct the waveform perfectly. Recognition models have the largest token resolution of 40 ms, barely smaller than the duration of a letter or phoneme to recognize. Token resolution is inversely proportional to the number of tokens for a given speech length, leading to differences in speed and memory usage across models. We will discuss the effect of it later. 

\subsection{Training and Evaluation}
\label{subsec:train_eval}

We trained all models with the same optimizer, learning rate, number of epochs, and data augmentation as the existing transformer receipts\footnote{
Sepformer:\url{https://github.com/speechbrain/speechbrain/blob/develop/recipes/WSJ0Mix/separation/hparams/sepformer.yaml}}
\footnote{
Conformer:\url{https://github.com/speechbrain/speechbrain/blob/develop/recipes/LibriSpeech/ASR/transformer/hparams/conformer_large.yaml}}
\footnote{
VALL-E:\url{https://github.com/lifeiteng/vall-e}
}. 
Mamba-TasNet (M, L) were trained with a cosine decay scheduler following \cite{dpmamba}, and Mamba-TasNet (L) was trained with a lower learning rate of $1.0e^{-4}$ for loss stability \cite{jamba}. Following convention, we trained Mamba-TasNet on WSJ0-2mix \cite{wsj0-2mix}, ConMamba on LibriSpeech \cite{librispeech}, and VALL-M on LibriTTS \cite{libritts}. Mamba-TasNet (M, L) and ConMamba (S, L) were trained on one NVIDIA L40. ConMamba (CTC) was trained on four A40s. VALL-M was trained on four L40s. Conformer (CTC) and VALL-E were reproduced on the same dataset and hardware. We evaluated separation models by the improvement of signal-to-distortion ratio (SDRi) and scale-invariant signal-to-noise ratio (SI-SDRi) \cite{sisnr}, ASR models by word error rate (WER) with or without an external pre-trained language model (with a small beam of 10 samples), and TTS models by comparative mean opinion score of speech naturalness and speaker similarity (CMOS-N and CMOS-S, with a scale of -6 to 6) compared to VALL-E.

\subsection{Performance}
\label{subsec:performance}
We report the performance of Mamba-TasNet, ConMamba, and VALL-M \& VALL-ME in Table.\ref{table:ss}, \ref{table:asr}, and \ref{table:tts}, compared to transformer models Sepformer, Conformer and VALL-E. For separation, we also compare with an earlier dual-path Mamba model \cite{dpmamba}. Mamba-TasNet (M) achieves a slightly higher performance than Sepformer in both SI-SNRi (+0.1dB) and SDRi (+0.2dB) despite with the same number of layers, Mamba-TasNet is 40\% smaller than Sepformer. Mamba-TasNet (M) is slightly worse than DPMamba (M), but when doubling the feature dimension from 256 to 512, Mamba-TasNet (L) outperforms both DPMamba (L) and QDPN, a transformer model more than three times larger. For ASR, the ConMamba encoder-only and the ConMamba encoder-transformer decoder model (S) outperform the Conformer encoder-only or the Conformer encoder-transformer decoder model (S) with a 1.0 and 0.5 lower WER on test-other without LM. In (L) size, the ConMamba encoder slightly loses to the ConMamba encoder, both with a transformer decoder. Fixing the ConMamba encoder, the Mamba decoder performs slightly worse than the transformer decoder in both (S) and (L) sizes. For TTS, we observe a similar result. The Mamba AR and Mamba NAR model VALL-M is worse than VALL-E, particularly regarding speech naturalness (-0.41). However, the transformer AR and Mamba NAR model VALL-ME receive a comparable (-0.01) score in naturalness and a higher (+0.20) score in similarity than VALL-E.
\\
In summary, the Mamba encoders perform on par or better than transformer encoders in separation, ConMamba encoders, and NAR LM, but the Mamba decoders perform slightly worse than transformer decoders as ASR decoders and AR LM.
\subsection{Efficiency}
\label{subsec:efficiency}
We benchmark the memory and speed of both Mamba and transformer models in all three tasks in Fig.\ref{fig:benchmark}. For separation and ASR, the memory and processing time are averaged for different input speech durations. For TTS, we plot the memory and time of all samples we benchmark since we cannot control the exact duration of the output speech.

\noindent \textbf{Memory} usage is smaller for Mamba-TasNet (M) than Sepformer in all speech durations. ConMamba and VALL-M have similar memory usage to Conformer and VALL-E until around 30 and 15 seconds, respectively. After, the memory for Conformer and VALL-E increases quadratically, while the memory for ConMamba and VALL-M increases linearly with a gentle slope. After 4 seconds, Mamba-TasNet (L) also consumes less memory than Sepformer, although the former is twice as large as the latter.

\noindent \textbf{Speed} follows a similar pattern as memory. Mamba-TasNet (M) is faster than Sepformer for all durations. Mamba-TasNet (L) catches up with Sepformer after 12 seconds. ConMamba and VALL-M are faster than Conformer and VALL-E after 60 and 15 seconds, respectively.

\section{Discussion and Limitation}
\label{sec:discussion}
We remark on the performance and efficiency of Mamba following the results. First, bidirectional Mamba challenges self-attention in performance. We get similar or better scores when replacing the transformer encoder layers in separation, Conformer encoder, and NAR LM with Mamba encoder layers. However, Mamba loses to cross- or masked attention for multimodal data: encoder acoustic features and decoder textual features for ASR and text prompts and speaker enrollment for TTS. The performance gap might be due to the gap in modeling power: Cross- or masked attention is not strictly causal in processing the keys or the unmasked part of the sequence, but Mamba in decoders processes the entire concatenated sequence, including the encoder's keys or the phoneme sequence, unidirectionally. Efficiency-wise, we observe Mamba is more efficient for long speech sequences than transformers. Referring to Table \ref{table:config}, we observe Mamba is most efficient for time-domain separation, where each token lasts 1 ms, and least for ASR, where each token lasts 40 ms. Figure \ref{fig:benchmark} shows ConMamba is not significantly more efficient than Conformer until 30 seconds for memory or 60 seconds for speed. Finally, we want to emphasize that this early study focuses on the vanilla transformer and Mamba architecture. We are not aiming for the best performance and efficiency with modified architecture. It is also possible that Mamba could be implemented more efficiently with better hardware and improved algorithms in the future. We hope our results and analysis can provide insights into the application of Mamba for different speech-processing scenarios.

\section{Conclusion}
This work presents Mamba-TasNet for separation, ConMamba for ASR, and VALL-M and VALL-ME for TTS. We demonstrate their comparable or higher performance and better efficiency in long speech than transformers. We observe this efficiency is inversely related to the resolution of a speech token. Mamba-TasNet is more efficient than Sepformer in all durations, while VALL-M and ConMamba are more efficient for durations of 15 seconds or longer. We conclude by discussing favorable use cases for Mamba in tasks requiring high speech resolution and its limitations, including joint modeling of multimodal data such as text and speech.

\section{ACKNOWLEDGMENT}
\label{sec:ack}
This work is funded by the National Institutes of Health (NIH- NIDCD) and a grant from Marie-Josee and Henry R. Kravis.

\bibliographystyle{IEEEbib}
\bibliography{strings,refs}

\end{document}